\documentclass{article}

\usepackage{graphicx}

\usepackage{color} 

\usepackage{amssymb}

\usepackage{times}

\begin{document}

\title{Efficient estimation of default correlation for multivariate jump-diffusion processes}

\author{Di Zhang and Roderick V.N. Melnik%
\thanks{Corresponding author. Email: rmelnik@wlu.ca, Tel.: +1 519 8841970, Fax: +1 519
8869351.}}

\date{Mathematical Modelling and Computational Sciences, \\
Wilfrid Laurier University, Waterloo, ON, Canada N2L 3C5}


\maketitle

\begin{abstract}
Evaluation of default correlation is an important task in credit
risk analysis. In many practical situations, it concerns the joint
defaults of several correlated firms, the task that is reducible to
a first passage time (FPT) problem. This task represents a great
challenge for jump-diffusion processes (JDP), where except for very
basic cases, there are no analytical solutions for such problems. In
this contribution, we generalize our previous fast Monte-Carlo
method (non-correlated jump-diffusion cases) for multivariate (and
correlated) jump-diffusion processes. This generalization allows
us, among other things, to evaluate the default events of several
correlated assets based on a set of empirical data. The developed
technique is an efficient tool for a number of other applications,
including credit risk and option pricing.
\end{abstract}


\vspace*{2ex}\noindent\textit{Keywords}: Default correlation, First passage time problem, Monte Carlo simulation

\vspace*{1ex}\noindent\textbf{JEL classification:} C15

\vspace*{1ex}\noindent\textbf{Mathematics Subject Classification
(2000):} 65C05, 60J10, 60J75

\section{Introduction}
\label{Introduction}

In the financial world, individual companies are usually linked
together via economic conditions, so default correlation, defined as
the risk of multiple companies' default together, has been an
important area of research in credit analysis with applications to
joint default, credit derivatives, asset pricing and risk
management.

Nevertheless, the development of efficient computational tools for
modeling default correlation is lagging behind its practical needs.
Currently, there are two dominant groups of theoretical models used
in default correlation. One is a reduced form model, such as in
\cite{Li:2000} that uses a Copula function to parameterize the
default correlation. Recently, Chen et al \cite{Chen:2003} have
translated the joint default probability into bivariate normal
probability function.

The second group of models for default correlation is a structural
form model. Zhou \cite{Zhou:2001:corr} and Hull et al
\cite{Hull:2001} are the first to incorporate default correlation
into the Black-Cox first passage structural model. They got similar
closed form solutions for two assets. However, their models cannot
easily be extended to more than two assets. Furthermore, the
developed models do not include jump-diffusion processes.

As demonstrated in \cite{Zhou:2001:jump}, jump risk becomes an
important factor in credit risk analysis. It is now widely
acknowledged that the standard Brownian motion model for market
behavior falls short of explaining empirical observations of market
returns and their underlying derivative prices \cite{Kou:2003}. The
(multivariate) jump-diffusion model that provides a convenient
framework for investigating default correlation with jumps becomes
more readily accepted in the financial world.

One of the major problems in default analysis is to determine when a
default will occur within a given time horizon, or in other words,
what the default rate is during such a time horizon. This problem is
reduced to a first passage time (FPT) problem that can be formalized
on a basis of a certain stochastic differential equations (SDE). It
concerns the estimation of the probability density of the time for a
random process to cross a specified threshold level. Unfortunately,
after including jumps, only special cases have analytical solutions.
For most practical cases, closed form solutions are unavailable and
we can only turn to the numerical procedures.

Monte Carlo simulation is such a candidate for solving the SDE,
arising in the context of FPT problem. In conventional Monte Carlo
methods, we need to discretize the time horizon into small enough
intervals in order to avoid discretization bias \cite{Platen:2003},
and we need to evaluate the processes at each discretized time which
is very time-consuming. Many researchers have contributed to the
field and enhanced the efficiency of Monte Carlo simulation. Atiya
and Metwally \cite{Atiya:2005,Metwally:2002} have developed a fast
Monte Carlo-type numerical method to solve the FPT problem.
Recently, we have reported an extension of this fast Monte
Carlo-type method in the context of multiple non-correlated
jump-diffusion processes \cite{Zhang:2006}.

In this contribution, we develop a methodology for solving the FPT
problem in the context of multivariate jump-diffusion processes. In
particular, we have generalized the reported fast Monte-Carlo method
(non-correlated jump-diffusion cases) for multivariate (and
correlated) jump-diffusion processes. The paper is organized in the
following way: Section \ref{Model} details the description of our
model. The algorithms are described in Section \ref{Methodology}.
Section \ref{Application} contains the simulation results and
discussions, followed by the conclusions given in Section
\ref{Conclusion}.

\section{Developing multi-dimensional model}
\label{Model}

\subsection{Default correlation}

In the market economy, individual companies are inevitably linked
together via dynamically changing economic conditions. Take two
firms A and B as an example, whose probabilities of default are
$P_A$ and $P_B$, respectively. Then the default correlation can be
defined as
\begin{equation}
  \rho_{AB}=\frac{P_{AB}-P_{A}P_{B}}{\sqrt{P_A(1-P_A)P_B(1-P_B)}},
\end{equation}
where $P_{AB}$ is the probability of joint default.

Now, we can write $P_{AB}$ as
$P_{AB}=P_{A}P_{B}+\rho_{AB}\sqrt{P_A(1-P_A)P_B(1-P_B)}$. If we
assume $P_{A}=P_{B}=p\ll 1$, then we have $P_{AB}\approx
p^2+\rho_{AB}p\approx\rho_{AB}p$. Thus it is apparent that the
default correlation $\rho_{AB}$ plays a key role in the joint
default with important implication in the field of credit analysis.
Zhou \cite{Zhou:2001:corr} and Hull et al \cite{Hull:2001} were the
first to incorporate default correlation into the Black-Cox first
passage structural model.

In \cite{Zhou:2001:corr} Zhou has proposed a first passage time
model to describe default correlations of two firms under the
``bivariate diffusion process'':
\begin{equation}
  \left[\begin{array}{cc}
   d\ln(V_1)\\
   d\ln(V_2)
   \end{array}\right]=
  \left[\begin{array}{cc}
   \mu_1\\
   \mu_2
   \end{array}\right]dt+
   \Omega\left[\begin{array}{cc}
   dz_1\\
   dz_2
   \end{array}\right],
  \label{zhou:two:assets}
\end{equation}
where $\mu_1$ and $\mu_2$ are constant drift terms, $z_1$ and $z_2$
are two independent standard Brownian motions, and $\Omega$ is a
constant $2\times 2$ matrix such that
\[
  \Omega\cdot\Omega'=\left[\begin{array}{cc}
    \sigma_1^2 & \rho\sigma_1\sigma_2 \\
    \rho\sigma_1\sigma_2 & \sigma_2^2\end{array}\right].
\]
The coefficient $\rho$ reflects the correlation between the
movements in the asset values of the two firms. Based on Eq.
(\ref{zhou:two:assets}), Zhou has deduced the closed form solution
of default correlations of two assets \cite{Zhou:2001:corr}.
However, none of the above models has included possible jumps.
Apparently, jumps have more significant importance in the default
correlation than often perceived. Indeed, simultaneous jumps may
enhance the chance of simultaneous defaults which increases the
correlation defaults.

\subsection{Multivariate jump-diffusion process}
\label{subsection:MJD}

Let us consider a more general case. In a complete probability space
$(\Omega,F,P)$ with information filtration $(F_t)$. Suppose that
$X_t=\ln(V_t)$ is a Markov process in some state space
$D\subset\mathbb{R}^n$, solving the stochastic differential equation
\cite{Duffie:2000}
\begin{equation}
  dX_t=\mu(X_t)dt+\sigma(X_t)dW_t+dZ_t,
  \label{AJD:DiffEq}
\end{equation}
where $W$ is an $(F_t)$-standard Brownian motion in $\mathbb{R}^n$;
$\mu:D\rightarrow\mathbb{R}^n$,
$\sigma:D\rightarrow\mathbb{R}^{n\times n}$, and $Z$ is a pure jump
process whose jumps have a fixed probability distribution $\nu$ on
$\mathbb{R}^n$ such that they arrive with intensity
$\{\lambda(X_t):t\ge 0\}$, for some
$\lambda:D\rightarrow[0,\infty)$. Under these conditions, the above
model is reduced to an affine model if \cite{Duffie:2000}:
\begin{eqnarray}
  & & \mu(X_t,t) = K_0 + K_1 X_t\nonumber\\
  & & (\sigma(X_t,t)\sigma(X_t,t)^\top)_{ij} = (H_0)_{ij}+(H_1)_{ij}X_j\nonumber\\
  & & \lambda(X_t) = l_0+l_1\cdot X_t,
\end{eqnarray}
where $K=(K_0,K_1)\in\mathbb{R}^n\times\mathbb{R}^{n\times n}$,
$H=(H_0,H_1)\in\mathbb{R}^{n\times n}\times\mathbb{R}^{n\times
n\times n}$, $l=(l_0,l_1)\in\mathbb{R}^n\times\mathbb{R}^{n\times
n}$.

If we assume that,
\begin{enumerate}
  \item Each $W_t$ in Eq. (\ref{AJD:DiffEq}) is independent;
  \item $K_1=0$, $H_1=0$ and $l_1=0$ in (4) that means the drift term, the diffusion
  process (Brownian motion) and the arrival intensity are independent with state
  vector $X_t$;
  \item The distribution of jump-size $Z_t$ is also independent with respect to $X_t$.
\end{enumerate}

In this scenario, we can rewrite Eq. (\ref{AJD:DiffEq}) as
\begin{equation}
  dX_t=\mu dt+\sigma dW_t+dZ_t,
  \label{JDP:multi}
\end{equation}
where
\[
  \mu = K_0,\;\sigma\sigma^\top = H_0,\;\lambda = l_0.
\]

At first sight, Eq. (\ref{JDP:multi}) is similar to Eq.
(\ref{zhou:two:assets}), but Eq. (\ref{JDP:multi}) is a much more
general model that can be applied to multiple firms, and where jumps
have been taken into account.

\subsection{First passage time distribution}
\label{subsection:FPTD}

Let us consider a firm $i$, as described by Eq. (\ref{JDP:multi}),
such that its state vector $X_i$ satisfies the following SDE:
\begin{equation}
  dX_i = \mu_{i}dt+\sum_{j}\sigma_{ij}dW_j+dZ_i = \mu_{i}dt+\sigma_{i}dW_i+dZ_i,
  \label{JDP:one}
\end{equation}
where $W_i$ is a standard Brownian motion and $\sigma_{i}$ is:
\[
  \sigma_{i}=\sqrt{\sum_{j}\sigma_{ij}^2}.
\]

We assume that in the interval $[0,T]$, total number of jumps for
firm i is $M_i$ times of jumps. Let the jump instants be $T_{1},
T_{2},\cdots,T_{M_i}$. Let $T_{0}=0$ and $T_{M_i+1}=T$. The
quantities $\tau_j$ equal to interjump times, which is
$T_{j}-T_{j-1}$. Let $X_{i}(T_{j}^{-})$ be the process value
immediately before the $j$th jump, and $X_{i}(T_{j}^{+})$ be the
process value immediately after the $j$th jump. The jump-size is
$X_{i}(T_{j}^{+})-X_{i}(T_{j}^{-})$, we can use such jump-sizes to
generate $X_{i}(T_{j}^{+})$ sequentially.

In a structural model, a firm defaults when the firm assets value
$V_t$ falls below a threshold level $D_V(t)$. In this contribution,
we use an exponential form, defining the threshold level by
$D_V(t)=\kappa\exp(\gamma t)$ as proposed in \cite{Zhou:2001:corr},
where $\gamma$ can be interpreted as the growth rate of firm's
liabilities. Coefficient $\kappa$, in front, captures the liability
structure of the firm, which is usually defined as a firm's
short-term liability plus 50\% of the firm's long-term liability. As
mentioned before, $X_t=\ln(V_t)$, then the threshold of $X_t$ is
$D(t)=\gamma t+\ln(\kappa)$.

Atiya et al \cite{Atiya:2005} have deduced a one-dimensional first
passage time distribution in time horizon $[0,T]$. In order to
evaluate multi-firms, we obtain multi-dimensional formulas and
reduce them to computable forms.

First, let us define $A_i(t)$ as the event that process crossed the
threshold level $D_i(t)$ for the first time in the interval
$[t,t+dt]$. Then we have
\begin{equation}
  g_{ij}(t)=p(A_i(t)\in dt|X_i(T_{j-1}^{+}),X_i(T_{j}^{-})).
\end{equation}

If we only consider one interval $[T_{j-1},T_{j}]$, we can obtain
\begin{eqnarray}
  g_{ij}(t) & = &
  \frac{X_i(T_{j-1}^{+})-D_{i}(t)}{2y_i\pi\sigma_{i}^{2}}(t-T_{j-1})^{-\frac{3}{2}}(T_{j}-t)^{-\frac{1}{2}}\nonumber\\
  & & *\exp\left(-\frac{[X_i(T_{j}^{-})-D_{i}(t)-\mu_{i}(T_{j}-t)]^{2}}{2(T_{j}-t)\sigma_{i}^{2}}\right)\nonumber\\
  & & *\exp\left(-\frac{[X_i(T_{j-1}^{+})-D_{i}(t)+\mu_{i}(t-T_{j-1})]^{2}}{2(t-T_{j-1})\sigma_{i}^{2}}\right),
  \label{FPTD:condition}
\end{eqnarray}
where
\[
  y_i=\frac{1}{\sigma_{i}\sqrt{2\pi\tau_{j}}}
    \exp\left(-\frac{[X_i(T_{j-1}^{+})-X_i(T_{j}^{-})+\mu_{i}\tau_{j}]^{2}}{2\tau_{j}\sigma_{i}^{2}}\right).
\]

After getting these results in one interval, we combine the results
to obtain the density for the whole interval $[0,T]$. Let $B(s)$ be
a Brownian bridge in the interval $[T_{j-1},T_{j}]$ with
$B(T_{j-1}^{+})=X_i(T_{j-1}^{+})$ and $B(T_{j}^{-})=X_i(T_{j}^{-})$.
Then the probability that the minimum of $B(s_i)$ is always above
the boundary level is
\begin{eqnarray}
  P_{ij} & = & P\left(\inf_{T_{j-1}\leq s_i\leq T_{j}}B(s_i)>D_{i}(t)|B(T_{j-1}^{+})=X_i(T_{j-1}^{+}),B(T_{j}^{-})=X_i(T_{j}^{-})\right)\nonumber\\
   & = & \left\{
     \begin{array}{ll}
       1-\exp\left(-\frac{2[X_i(T_{j-1}^{+})-D_{i}(t)][X_i(T_{j}^{-})-D_{i}(t)]}{\tau_{j}\sigma_{i}^{2}}\right), & \mathrm{if}\;X_i(T_{j}^{-})>D_{i}(t),\\
       0, & \mathrm{otherwise}.
     \end{array}\right.
   \label{BM:default}
\end{eqnarray}

This implies that $B(s_i)$ is below the threshold level, which means
the default happens or already happened, and its probability is
$1-P_{ij}$. Let $L(s_i)\equiv L_i$ denote the index of the interjump
period in which the time $s_i$ falls in $[T_{L_i-1},T_{L_i}]$. Also,
let $I_i$ represent the index of the first jump, which happened in
the simulated jump instant:
\begin{eqnarray}
  I_i & = & \min(j:X_i(T_{k}^{-})>D_{i}(t);k=1,\ldots,j,\;\mathrm{and}\nonumber\\
    &   & \;\;\;\;\;\;\;\;\;\;\;\:X_i(T_{k}^{+})>D_{i}(t);k=1,\ldots,j-1,\;\mathrm{and}
          \;X_i(T_{j}^{+})\leq D_{i}(t)).
  \label{index_first_jump}
\end{eqnarray}

If no such $I_i$ exists, then we set $I_i=0$.

By combining Eq. (\ref{FPTD:condition}), (\ref{BM:default}) and
(\ref{index_first_jump}), we get the probability of $X_i$ crossing
the boundary level in the whole interval $[0,T]$ is
\begin{eqnarray}
  & & P(A_i(s_i)\in ds|X_i(T_{j-1}^{+}),X_i(T_{j}^{-}),j=1,\ldots,M_{i}+1)\nonumber\\
   & = & \left\{
     \begin{array}{ll}
       g_{iL_i}(s_i)\prod_{k=1}^{L_i-1}P_{ik} & \mathrm{if}\;L_i<I_i\;\mathrm{or}\;I_i=0,\\
       g_{iL_i}(s_i)\prod_{k=1}^{L_i-1}P_{ik}+\prod_{k=1}^{L_i}P_{ik}\delta(s_i-T_{I_i}) & \mathrm{if}\;L_i=I_i,\\
       0 & \mathrm{if}\;L_i>I_i,
     \end{array}\right.
\end{eqnarray}
where $\delta$ is the Dirac's delta function.

\subsection{The kernel estimator}
\label{subsection:estimation}

For each firm, after generating a series of first passage times
$s_i$, we use a kernel density estimator with Gaussian kernel to
estimate the first passage time density (FPTD) $f$. As described in
\cite{Atiya:2005}, the kernel density estimator is based on
centering a kernel function of a bandwidth as follows:
\begin{equation}
  \widehat{f}=\frac{1}{N}\sum_{i=1}^{N}K(h,t-s_{i}),
\end{equation}
where
\[
  K(h,t-s_{i})=\frac{1}{\sqrt{\pi/2}h}\exp\left(-\frac{(t-s_{i})^{2}}{h^2/2}\right).
\]

The optimal bandwidth in the kernel function $K$ can be calculated
as \cite{Silverman:1986}:
\begin{equation}
  h_{opt}=\left(2N\sqrt{\pi}\int_{-\infty}^{\infty}(f_{t}'')^{2}dt\right)^{-0.2},
  \label{estamate:hopt}
\end{equation}
where $N$ is the number of generated points and $f_{t}$ is the true
density. Here we use the approximation for the distribution, as a
gamma distribution as proposed in \cite{Atiya:2005}:
\begin{equation}
  f_{t}=\frac{\alpha^{\beta}}{\Gamma(\beta)}t^{\beta-1}\exp(-\alpha t).
\end{equation}

So the integral in Eq. (\ref{estamate:hopt}) becomes,
\begin{equation}
  \int_{0}^{\infty}(f_{t}'')^{2}dt=
    \sum_{i=1}^{5}\frac{W_{i}\alpha_{i}\Gamma(2\beta-i)}{2^{(2\beta-i)}(\Gamma(\beta))^{2}},
  \label{Eq:hopt2}
\end{equation}
where
\[
  W_{1}=A^{2},\;\;W_{2}=2AB,\;\;W_{3}=B^{2}+2AC,\;\;W_{4}=2BC,\;\;W_{5}=C^{2},
\]
and
\[
  A=\alpha^{2},\;\;B=-2\alpha(\beta-1),\;\;C=(\beta-1)(\beta-2).
\]

From Eq. (\ref{Eq:hopt2}), follows that in order to get a nonzero
bandwidth, we have constraint $\beta$ to be at least equal to 3.

After obtaining the estimated first passage time density
$\widehat{f}$, the cumulative default rates can be written as,
\begin{equation}
  P_i(t)=\int_{0}^{t}\widehat{f}_{i}(\tau)d\tau.
\end{equation}

\section{Methodology of the solution}
\label{Methodology}

In Section \ref{Model}, we have reduced the solution of the original
problem to a multivariate jump-diffusion model as described in Eq.
(\ref{JDP:one}). The first passage time distribution was obtained in
Section \ref{subsection:FPTD}. As we have already noted, once jumps
are included in the process, only for very basic applications closed
form solutions are available \cite{Kou:2003,Blake:1973}, which in
most practically interesting cases we have to resort to the
numerical procedures.

\begin{figure}[hbtp]
  \centering
 \includegraphics[width=8cm, angle=270]{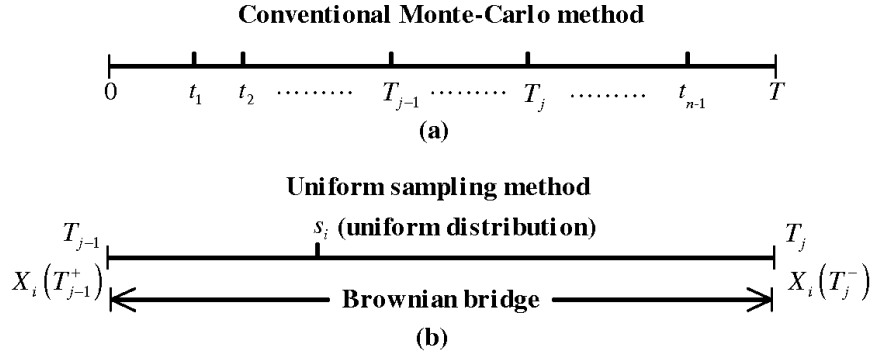}
  \caption{Schematic diagram of (a) conventional Monte Carlo and
           (b) uniform sampling method.}
  \label{Fig:Method}
\end{figure}

Let us recall the conventional Monte-Carlo procedure in application
to the analysis of the evaluation of firm $X_i$ within the time
horizon $[0,T]$. We divide the time horizon into $n$ small intervals
$[0,t_1]$, $[t_1,t_2]$, $\cdots$, $[t_{n-1},T]$ as displayed in Fig.
\ref{Fig:Method}(a). In each Monte Carlo run, we need to calculate
the value of $X_i$ at each discretized time $t$. As usual in order
to exclude discretization bias, the number $n$ must be large. This
procedure exhibits substantial computational difficulties when
applied to jump-diffusion process.

Indeed, for a typical jump-diffusion process, as shown in Fig.
\ref{Fig:Method}(a), let $T_{j-1}$ and $T_j$ be any successive jump
instants, as described above. Then, in the conventional Monte Carlo
method, although there is no jump occurring in the interval
$[T_{j-1},T_j]$, yet we need to evaluate $X_i$ at each discretized
time $t$ in $[T_{j-1},T_j]$. This very time-consuming procedure
results in serious shortcoming of the conventional Monte Carlo
methodology.

To remedy the situation, two modifications of the conventional
procedure were recently proposed by Atiya and Metwally
\cite{Atiya:2005,Metwally:2002} that allow us a potential speed-up
of the conventional methodology in 10-30 times. One of the
modifications, the uniform sampling method, involves sampling
method. The other, inverse Gaussian density sampling, is based on
the inverse Gaussian density method for sampling. Both methodologies
were developed for the univariate case.

In this article, we focus on the further development of uniform
sampling (UNIF) method and extend it to multivariate jump-diffusion
processes. The major improvement of the uniform sampling method is
based on the fact that it only evaluates $X_i$ at generated jump
instants while between each two jumps the process is a Brownian
bridge (see Fig. \ref{Fig:Method}(b)). Hence we just consider the
probability of $X_i$ defaults in $(T_{j-1},T_j)$ instead of
evaluating $X_i$ at each discretized time $t$. More precisely, in
the uniform sampling method, we assume that the values of
$X_i(T_{j-1}^+)$ and $X_i(T_j^-)$ are known as two end points of the
Brownian bridge, we generate a variable $s_i$ with uniform
distribution and by using Eq. (\ref{BM:default}) we verify whether
$X_i(s_i)$ is smaller than the threshold level. If it is, then we
have successfully generated a first passage time $s_i$ and can
neglect the other intervals and perform another Monte Carlo run.

In order to implement the UNIF method for our multivariate model as
described in Eq. (\ref{JDP:one}), we need to consider several points
as follows:
\begin{enumerate}
  \item Here, we focus on the  firms rated in the same way, that is we assume that
the arrival rate $\lambda$ for the Poisson jump process and the
distribution of $(T_j-T_{j-1})$ are the same for each firm. As for
the jump-size, we generate them by a given distribution, which can
be different for different firms to reflect specifics of the jump
process for each firm.
  \item We exemplify our description by considering an
    exponential distribution (mean value $\mu_T$) for
$(T_j-T_{j-1})$ and a normal distribution (mean value $\mu_J$ and
standard deviation $\sigma_J$) for the jump-size. We can use any
other distribution when appropriate.
  \item An array \texttt{IsDefault} (whose size is the number of firms denoted by
$N_{\mathrm{firm}}$) was used to indicate whether firm $i$ has
defaulted in this Monte Carlo run. If the firm defaults, then we set
\texttt{IsDefault}$(i)=1$, and will not evaluate it during this
Monte Carlo run.
  \item Most importantly, in order to reflect the correlations of
  multiple firms, we need to generate correlated $s_i$. As deduced
  in \cite{Iyengar:1985,Zhou:2001:corr}, the joint probability of
  firm $i$ to default before $s_i$ and firm $j$ to default before
  $s_j$ satisfies the bivariate inverse Gaussian distribution. We
  approximate the correlation between $s_i$ and $s_j$, denoted
  further by $\rho_{s_i,s_j}$ as the default correlation of  firm $i$
  and $j$. Then, we employ the sum-of-uniforms method
  \cite{Chen:2005} to generate correlated uniform numbers $s_i$.
\end{enumerate}

Our algorithm for multivariate jump-diffusion processes can be
described as follows. It is an extension of the one-dimensional
case, described in \cite{Atiya:2005,Metwally:2002}.

Consider $N_{\mathrm{firm}}$ firms in the interval $[0,T]$. First,
we generate the jump instant $T_{j}$ by generating interjump times
$(T_{j}-T_{j-1})$, and set all the \texttt{IsDefault}$(i)=0$ to
indicate that no firm defaults at first.

From Fig. \ref{Fig:Method}(b) and Eq. (\ref{JDP:one}), we can
conclude that,
\begin{enumerate}
  \item If no jump occurs, as described in Eq. (\ref{JDP:one}), the
interjump size $(X_i(T_{j}^{-})-X_i(T_{j-1}^{+}))$ follows a normal
distribution of mean $\mu_i(T_{j}-T_{j-1})$ and standard deviation
$\sigma_i\sqrt{T_{j}-T_{j-1}}$. We get
\begin{eqnarray*}
X_i(T_{j}^{-})&\sim&X_i(T_{j-1}^{+})+\mu_i(T_{j}-T_{j-1})+
\sigma_{i}N(0,T_{j}-T_{j-1})\\
&\sim&X_i(T_{j-1}^{+})+\mu_i(T_{j}-T_{j-1})+
\sum_{k=1}^{N_{\mathrm{firm}}}\sigma_{ik}N(0,T_{j}-T_{j-1}),
\end{eqnarray*}
where the initial state is $X_i(0)=X_i(T_{0}^{+})$.
  \item If jump occurs, we simulate the jump-size by a normal distribution
or other distribution when appropriate, and compute the postjump
value:
\[
  X_i(T_{j}^{+})=X_i(T_{j}^{-})+Z_i(T_{j}).
\]
\end{enumerate}

After generating beforejump and postjump values $X_i(T_{j}^{-})$ and
$X_i(T_{j}^{+})$. As before, $j=1,...,M$ where $M$ is the number of
jumps for all the firms, we can compute $P_{ij}$ according to Eq.
(\ref{BM:default}). To recur the first passage time density (FPTD)
$f_i(t)$, we have to consider three possible cases that may occur
for each non-default firm $i$:
\begin{enumerate}
  \item \textbf{First passage happens inside the interval.} We know if
$X_i(T_{j-1}^{+})>D_i(T_{j-1})$ and $X_i(T_{j}^{-})<D_i(T_{j})$,
then the first passage happened in the time interval
$[T_{j-1},T_{j}]$. To evaluate when the first passage happened, we
introduce a new viable $b_{ij}$, as
$b_{ij}=\frac{T_{j}-T_{j-1}}{1-P_{ij}}$, so that
$T_{j-1}+b_{ij}=\frac{T_{j}-P_{ij}T_{j-1}}{1-P_{ij}}$. By using the
sum-of-uniforms method, we generate several correlated uniform
numbers $s_i$ in the interval of $[T_{j-1},T_{j-1}+b_{ij}]$, and if
$s_i$ also belongs to interval $[T_{j-1},T_{j}]$, then the first
passage time occurred in this interval. We set
\texttt{IsDefault}$(i)=1$ to indicate firm $i$ that has defaulted.
Then, we can compute the conditional boundary crossing density
$g_{ij}(s_i)$ according to Eq. (\ref{FPTD:condition}). To get the
density of the entire interval $[0,T]$, we  use
$\widehat{f}_{i,n}(t)=\left(\frac{T_{j}-T_{j-1}}{1-P_{ij}}\right)g_{ij}(s_i)*K(h_{opt},t-s_i)$,
where $n$ is the iteration number of the Monte Carlo cycle.
  \item \textbf{First passage does not happen in the current interval.} If $s_i$
doesn't belong to interval $[T_{j-1},T_{j}]$, then the first passage
time has not yet occurred in this interval.
  \item \textbf{First passage happens at the right boundary of interval.} If
$X_i(T_{j}^{+})<D_i(T_{j})$ and $X_i(T_{j}^{-})>D_i(T_{j})$ (see
Eq. (\ref{index_first_jump})), then  $T_{I_i}$  is the first passage
time and $I_i=j$. We evaluate the density function using kernel
function $\widehat{f}_{i,n}(t)=K(h_{opt},t-T_{I_i})$, and set
\texttt{IsDefault}$(i)=1$.
\end{enumerate}

Next, we increase $j$ and examine the next interval and analyze the
above three cases for each non-default firm again. After running $N$
times the Monte Carlo cycle, we get the FPTD of firm $i$ as
$\widehat{f}_{i}(t)=\frac{1}{N}\sum_{n=1}^{N}\widehat{f}_{i,n}(t)$,
as well as the cumulative default rates
$P_{i}(t)=\int_{0}^{t}\widehat{f}_{i}(\tau)d\tau$.

\section{Applications and discussion}
\label{Application}

In this section, we will demonstrate how our model describes the
default correlations of the firms, rated in the same way, via
studying the historical data. We provide details on the calibration
of the models we apply for this description.

\subsection{Default rates}
\begin{figure}[hbtp]
  \centering
  \includegraphics[width=14cm]{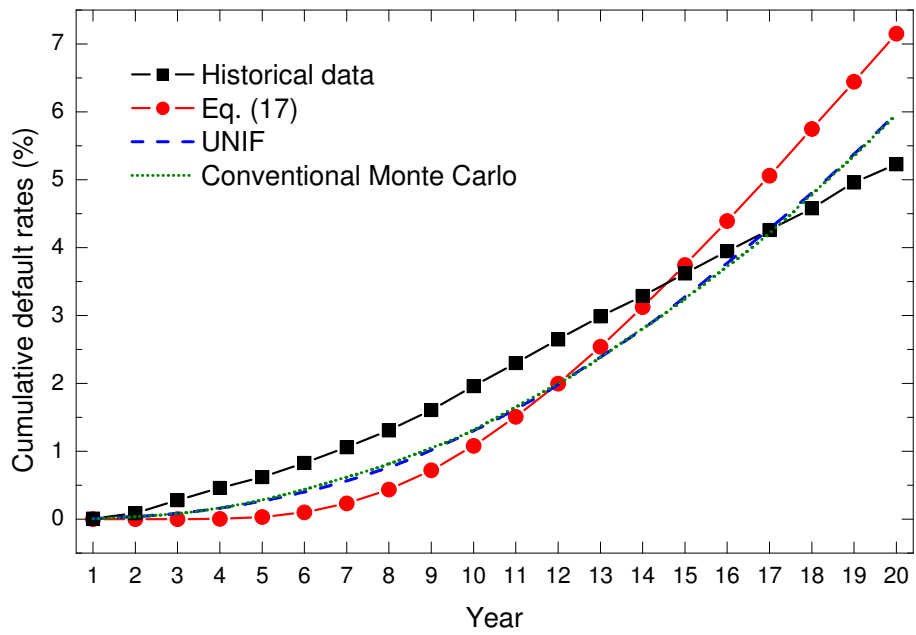}
  \caption{
Historical, theoretical and simulated cumulative default rates. The
theoretical value is calculated by using Eq.
(\ref{default:zhou:model}). All the simulations were performed with
Monte Carlo runs $N=100,000$, besides, for conventional Monte Carlo
method, the discretization size of time horizon is $\Delta=0.005$.}
  \label{Fig:default:one}
\end{figure}

In Fig. \ref{Fig:default:one}, the black  line of square is a set of
historical default data of A-rated firm\footnote{A-rated firm stands
for a specific kind of firm following the Moody's Investors
Service's definition.} taken from \cite{Zhou:2001:corr}.

First, if we do not consider jumps, as assumed in
\cite{{Zhou:2001:corr}}, the firm defaults at time $t$ with
probability:
\begin{equation}
  P_i(t)=2\cdot N\left(-\frac{X_i(0)-\ln(\kappa_i)}{\sigma_i\sqrt{t}}\right)%
        =2\cdot N\left(-\frac{Z_i}{\sqrt{t}}\right),
  \label{default:zhou:model}
\end{equation}
where
\[
  Z_i\equiv\frac{X_i(0)-\ln(\kappa_i)}{\sigma_i}
  \label{Zi:zhou}
\]
is the standardized distance of firm $i$ to its default point and
$N(\cdot)$ denotes the cumulative probability distribution function
for a standard normal variable.

If historical default rates are given, we can estimate $Z_i$ as
follows:
\begin{equation}
  Z_i=\mathrm{arg}\min_{Z_i}\sum_{t}\left(\frac{P_i(Z_i,t)}{t}-\frac{\widetilde{A}_i(t)}{t}\right)^2,
\end{equation}
where $P_i(Z_i,t)$ are the theoretical default probabilities (as
determined by Eq. (\ref{default:zhou:model})) and
$\widetilde{A}_i(t)$ are the historical default rates. For the
A-rated firm considered here, the optimized $Z_i$ value was
evaluated in  \cite{Zhou:2001:corr} as  $8.06$. By feeding the
optimized $Z_i$-value into Eq. (\ref{default:zhou:model}), we get
the theoretical cumulative default rates without jumps, given in
Fig. \ref{Fig:default:one} by the line of circles.

Now, let us consider the UNIF method, developed in Section
\ref{subsection:estimation} and \ref{Methodology}. First, the
developed Monte Carlo simulation allows us to obtain the estimated
density $\widehat{f}_{i}(t)$ by using kernel estimator method. We
get also the default rate $P_{i}(t)$ for firm $i$.

Then we minimize the difference between our model and the historical
default data to obtain the optimized parameters in our model:
\begin{equation}
  \mathrm{argmin}\left(\sum_{i}\sqrt{\sum_{t_j}\left(
    \frac{P_{i}(t_j)-\widetilde{A}_i(t_j)}{t_j}\right)^2}\right).
  \label{Eq:calibration:default}
\end{equation}

For convenience, we reduce the number of optimizing parameters by:
\begin{enumerate}
  \item Setting $X(0)=2$ and $\ln(\kappa)=0$.
  \item Setting the growth rate $\gamma$ of debt value equivalent to the
growth rate $\mu$ of the firm's value \cite{Zhou:2001:corr}, so the
default of firm is non-sensitive to $\mu$. Setting $\mu=-0.001$ in
our computation reported next.
  \item The interjump times $(T_j-T_{j-1})$ satisfy an exponential
  distribution with mean value equals to 1.
  \item The arrival rate for jumps satisfies the Poisson distribution with
intensity parameter $\lambda$, where the jump size is a normal
distribution $Z_t\sim N(\mu_{Z},\sigma_{Z})$.
\end{enumerate}

As a result, we only need to optimize $\sigma$, $\lambda$,
$\mu_{Z}$, $\sigma_{Z}$ for this firm, This is done by minimizing the
differences between our simulated default rates and the given
historical data. The minimization was performed by using
quasi-Newton procedure implemented as a Scilab program.

The optimized parameters for the A-rated firm are
$\sigma=0.09000984$, $\lambda=0.10001559$, $\mu_{Z}=-0.20003641$,
and $\sigma_{Z}=0.50000485$. Then, by using these optimized
parameters, we carried out a final simulation with Monte Carlo runs
$N=100,000$. The simulated cumulative default rates by using the UNIF
method are shown in Fig. \ref{Fig:default:one} by the  dash line.
For comparison, we have carried out the conventional Monte Carlo
simulation with the same optimized parameters. The resulting
simulated default rates are displayed by  dotted line in Fig.
\ref{Fig:default:one}. All the simulations reported here were
carried out on a 2.4 GHz AMD Opteron(tm) Processor. The optimal
bandwidth and CPU time are given in Table \ref{Tab:hopt}.

\begin{table}[hbtp]
  \centering
  \caption{The optimal bandwidth $h_{opt}$, and CPU time per Monte Carlo run
of the simulations. All the simulations were performed with Monte
Carlo runs $N=100,000$;  the discretization size of time horizon for
the conventional Monte Carlo method was $\Delta=0.005$.}
  \label{Tab:hopt}
  \begin{tabular}{lcc}
    \hline
    & Optimal bandwidth & CPU time per Monte Carlo run\\
    \hline
    Conventional Monte Carlo & 0.891077 & 0.119668\\
    UNIF & 0.655522 & 0.000621\\
    \hline
  \end{tabular}
\end{table}

From Fig. \ref{Fig:default:one}, we can conclude that our
simulations give similar results to the theoretically predicted by
Eq. (\ref{default:zhou:model}), and exceed them for short time
horizon. The UNIF method gives exactly the same default curve as the
conventional Monte Carlo method, but the former outperforms the
latter substantially in terms of computational time. The UNIF
methodology is much faster compared to the conventional method and
is extremely useful in practical application.

\subsection{Default correlations}
Our final example concerns with the default correlation of two
A-rated firms (A,A). In Table \ref{Simulate:Corr} we provide the
information on the default correlation of firms (A,A) for one-,
two-, five- and ten-year. The values in the 2nd column were
calculated by using the closed form solution derived in
\cite{Zhou:2001:corr}.

\begin{table}[hbtp]
  \centering
  \caption{Theoretical and simulated default correlations (\%) of firms (A,A).
The simulations were performed with Monte Carlo runs $N=100,000$.}
  \begin{tabular}{ccc}
    \hline
    Year & Ref. \cite{Zhou:2001:corr} & UNIF \\
    \hline
     1  &  0.00 &  0.00\\
     2  &  0.02 &  2.47\\
     5  &  1.65 &  6.58\\
    10  &  7.75 &  9.28\\
    \hline
  \end{tabular}
  \label{Simulate:Corr}
\end{table}

In order to implement the UNIF method, we use assumptions,
similar to the ones before, in order to reduce the number of
optimized parameters:
\begin{enumerate}
  \item Setting $X(0)=2$ and $\ln(\kappa)=0$ for all firms.
  \item Setting $\gamma=\mu$ and $\mu=-0.001$ for all firms.
  \item Since we are considering two same rated firms (A,A), we choose $\sigma$
 as:
\begin{equation}
  \sigma=\left[
  \begin{tabular}{cc}
    $\sigma_{11}$ & $\sigma_{12}$\\
    $\sigma_{21}$ & $\sigma_{22}$
  \end{tabular}\right],
\end{equation}
where $\sigma\sigma^\top=H_0$ such that
\begin{equation}
  \sigma\sigma^\top=H_0=\left[
  \begin{tabular}{cc}
    $\sigma_1^2$ & $\rho_{12}\sigma_1\sigma_2$ \\
    $\rho_{12}\sigma_1\sigma_2$ & $\sigma_2^2$
  \end{tabular}\right],
\end{equation}
and
\begin{equation}
  \left\{
  \begin{tabular}{l}
    $\sigma_1^2=\sigma_{11}^2+\sigma_{12}^2$\\
    $\sigma_2^2=\sigma_{21}^2+\sigma_{22}^2$\\
    $\rho_{12}=\displaystyle\frac{\sigma_{11}\sigma_{21}+\sigma_{12}\sigma_{22}}{\sigma_1\sigma_2}$
  \end{tabular}\right.,
  \label{Eq:Brownian:corr}
\end{equation}
In (\ref{Eq:Brownian:corr}), $\rho_{12}$ reflects the
correlation of diffusion parts of the state vectors of the two
firms.
  \item The arrival rate for jumps satisfies the Poisson
distribution with intensity parameter $\lambda$ for all firms, and
we use the parameters optimized based on a single A-rated firm,
i.e., $\lambda=0.10001559$ for all the firms.
  \item As before, we generate the same interjump times $(T_j-T_{j-1})$ that
  satisfies an
exponential distribution with mean value equals 1 for all firms.
Furthermore, the jump size is a normal distribution $Z_t\sim
N(\mu_{Z_i},\sigma_{Z_i})$, and we use the parameters optimized from
a single A-rated firm, i.e., $\mu_{Z_i}=-0.20003641$ and
$\sigma_{Z_i}=0.50000485$ for all the firms.
\end{enumerate}
As a result, there are only 4 parameters left to optimize:
$\sigma_{11}$, $\sigma_{12}$, $\sigma_{21}$ and $\sigma_{22}$.  The
optimization was carried out by using the quasi-Newton procedure
implemented as a Scilab program. The resulting  optimized parameters
are $\sigma_{11}=0.06963755$, $\sigma_{12}=0.02993134$,
$\sigma_{21}=0.03387809$ and $\sigma_{22}=0.06691001$. We can easily
get $\sigma_1=0.0757976$, $\sigma_2=0.0749978$ and
$\rho_{12}=0.7673104$. The parameter $\rho_{12}$ represents the
correlation between  diffusion parts of the state vectors of
 two firms.

The simulated default correlations are displayed in the 3rd column
of Table \ref{Simulate:Corr}. Observe that, the UNIF method gives a
little larger default correlation compared to the theoretical
predicted by Eq. (\ref{default:zhou:model}). This is mainly because
our optimized $\rho_{12}$ is larger than $0.4$ used in
\cite{Zhou:2001:corr}, and we have used the same interjump times
$(T_j-T_{j-1})$ for all the firms. Nevertheless, the UNIF method
gives the correct default correlation trend, as the default
correlation becomes larger with increasing time.

\section{Conclusion}
\label{Conclusion}

We analyzed the first passage time problem in the context of
multivariate and correlated jump-diffusion processes by extending
the fast Monte Carlo-type numerical method -- the UNIF method -- to
the multivariate case. We provided an application example of
simulating default correlations confirming the validity of our model
and the developed algorithm. Finally, we note that the developed
methodology provides an efficient tool for further practical
applications such as in credit analysis and barrier option pricing.

\section*{Acknowledgement}
  This work was supported by NSERC.




\end{document}